\documentclass[11pt, a4paper]{article}
\pdfoutput=1
\usepackage{jcappub}

\usepackage{booktabs}
\usepackage{multirow}
\usepackage{amssymb}
\usepackage[export]{adjustbox}
\usepackage{floatrow}
\newfloatcommand{capbtabbox}{table}[][3.5in]
\newfloatcommand{capbfigbox}{figure}[][2.3in]
\usepackage{blindtext}
\usepackage{amsmath}
\usepackage{booktabs}
\usepackage{aas_macros}

\def\lsim{\mathrel{\raise.3ex\hbox{$<$\kern-.75em\lower1ex\hbox{$\sim$}}}}
\def\gsim{\mathrel{\raise.3ex\hbox{$>$\kern-.75em\lower1ex\hbox{$\sim$}}}}

\begin{document}

\hspace*{110mm}{\large \tt FERMILAB-PUB-16-291-A}

\vskip 0.2in

\title{The Density of Dark Matter in the Galactic Bulge and Implications for Indirect Detection}

\author{Dan Hooper}\note{ORCID: http://orcid.org/0000-0001-8837-4127}
\emailAdd{dhooper@fnal.gov}

\affiliation{Fermi National Accelerator Laboratory, Center for Particle
Astrophysics, Batavia, IL 60510}
\affiliation{University of Chicago, Department of Astronomy and Astrophysics, Chicago, IL 60637}
\affiliation{University of Chicago, Kavli Institute for Cosmological Physics, Chicago, IL 60637}

\abstract{A recent study, making use of the number of horizontal branch stars observed in infrared photometric surveys and kinematic measurements of M-giant stars from the BRAVA survey, combined with N-body simulations of stellar populations, has presented a new determination of the dark matter mass within the bulge-bar region of the Milky Way. That study constrains the total mass within the $\pm 2.2 \times \pm 1.4 \times \pm 1.2$ kpc volume of the bulge-bar region to be ($1.84 \pm 0.07) \times 10^{10} \, M_{\odot}$, of which 9-30\% is made up of dark matter. Here, we use this result to constrain the the Milky Way's dark matter density profile, and discuss the implications for indirect dark matter searches. Although uncertainties remain significant, these results favor dark matter distributions with a cusped density profile. For example, for a scale radius of 20 kpc and a local dark matter density of 0.4 GeV/cm$^3$, density profiles with an inner slope of 0.69 to 1.40 are favored, approximately centered around the standard NFW value. In contrast, profiles with large flat-density cores are disfavored by this information. 
}

\maketitle

\section{Introduction}

The density of dark matter in the central volume of the Milky Way can strongly impact the prospects for indirect detection, as well as interpretations of the results of indirect detection experiments. This is particularly true for gamma-ray observations of the Galactic Center. The flux of dark matter annihilation products predicted from the innermost degree around the Galactic Center (corresponding to approximately the angular resolution of Fermi's Large Area Telescope) can vary by several orders of magnitude, depending on the halo profile that is adopted~\cite{Hooper:2012sr,Gomez-Vargas:2013bea}. For ground-based gamma-ray telescopes, with much greater angular resolution, the dependance on the halo profile's inner slope can be even more significant~\cite{Abramowski:2011hc,Aharonian:2006wh,Silverwood:2014yza,Pierre:2014tra}.

Although many groups have presented dynamical evidence in support of dark matter's presence in the Milky Way~\cite{Weber:2009pt,2010JCAP...08..004C,2011JCAP...11..029I,Bovy:2012tw,Garbari:2012ff,Bovy:2013raa,Read:2014qva,Iocco:2015xga,Pato:2015dua}, these measurements have until recently had little to say about the density of dark matter in the Galaxy's innermost kiloparsecs. In this short paper, we make use of the results of Portail {\it et al.}, who last year published a determination of the stellar mass and dark matter mass for the bulge-bar system of the Milky Way~\cite{2015MNRAS.448..713P}. In that study, the authors made use of the recently measured three-dimensional distribution of horizontal branch stars from infrared photometric surveys~\cite{2013MNRAS.435.1874W,2015MNRAS.450.4050W} and kinematic data for a large number of M-giant stars from the BRAVA survey~\cite{2008ApJ...688.1060H,2012AJ....143...57K}, combined with N-body simulations of stellar populations, to create dynamical models of the Galactic Bulge and Bar. From this information, they were able to constrain the total mass within the volume of the Bulge (taken to be a $\pm 2.2 \times \pm 1.4 \times \pm 1.2$ kpc box centered on the dynamical center of the Milky Way, corresponding to a total volume of 29.6 kpc$^3$) to be ($1.84 \pm 0.07) \times 10^{10} \, M_{\odot}$, including both statistical and systematic errors. While this result is consistent with previous determinations~\cite{1992ApJ...387..181K,2003MNRAS.340..949B,1995ApJ...449..521H,1995ApJ...444L..89B}, it is of unprecedented accuracy and appears to be quite robust to systematic uncertainties. The spatial morphology of stars in the Galaxy's bulge and bar is also now known in some detail, due in large part to modern photometric surveys~\cite{2015MNRAS.447.1535N}. Furthermore, Portail {\it et al.} find that the stellar mass within this volume, based on the observed number of red giants~\cite{2013ApJ...769...88N}, is consistently smaller than the total mass, allowing one to estimate the fraction of the Bulge's mass that consists of dark matter. Across the range of models presented by Portal {\it et al.}, the content of dark matter within the volume of the bulge-bar varies between $1.7\times 10^9 \, M_{\odot}$ and $5.3\times 10^9 \, M_{\odot}$, corresponding to 9-30\% of the total mass in the region. The largest source of uncertainty in this result is associated with the initial mass function, which is used to relate the number of red giants to a total stellar mass.

In this paper, we discuss the Milky Way's dark matter halo profile in light of this information, and assess the implications of this result for the indirect detection of dark matter in the Galactic Center and Inner Galaxy.  We find that the results of Portail {\it et al.} support the existence of a density cusp, and disfavor the presence of a large flat density core in the Milky Way's dark matter halo profile.

\section{Constraining The Milky Way's Dark Matter Density Profile}

In this study, we parameterize the Milky Way's dark matter halo distribution using the following generalized Navarro-Frenk-White (NFW) profile~\cite{Navarro:1995iw,Navarro:1996gj}:
\begin{equation}
\rho(r) \propto \frac{1}{r^{\gamma} \, [1+(r/R_s)]^{3-\gamma}}.
\end{equation}
In this expression, $\gamma$ is the inner slope of the profile ($\gamma=1$ corresponds to the case of a standard NFW profile), and $R_s$ is the scale radius. The flux of gamma rays or neutrinos generated by dark matter annihilation from a given direction of the sky is proportional to what is known as the $J$-factor~\cite{Bergstrom:1997fj}, which is given by:
\begin{equation}
J(\psi) = \int_{\rm los} \rho^2(l,\psi) \, dl(\psi),
\end{equation}
where $\psi$ is the angle observed and the integral is performed over the line-of-sight (los). In practice, one generally integrates the $J$-factor over a solid angle representing the field-of-view under consideration. 

\begin{figure}
\includegraphics[keepaspectratio,width=0.7\textwidth]{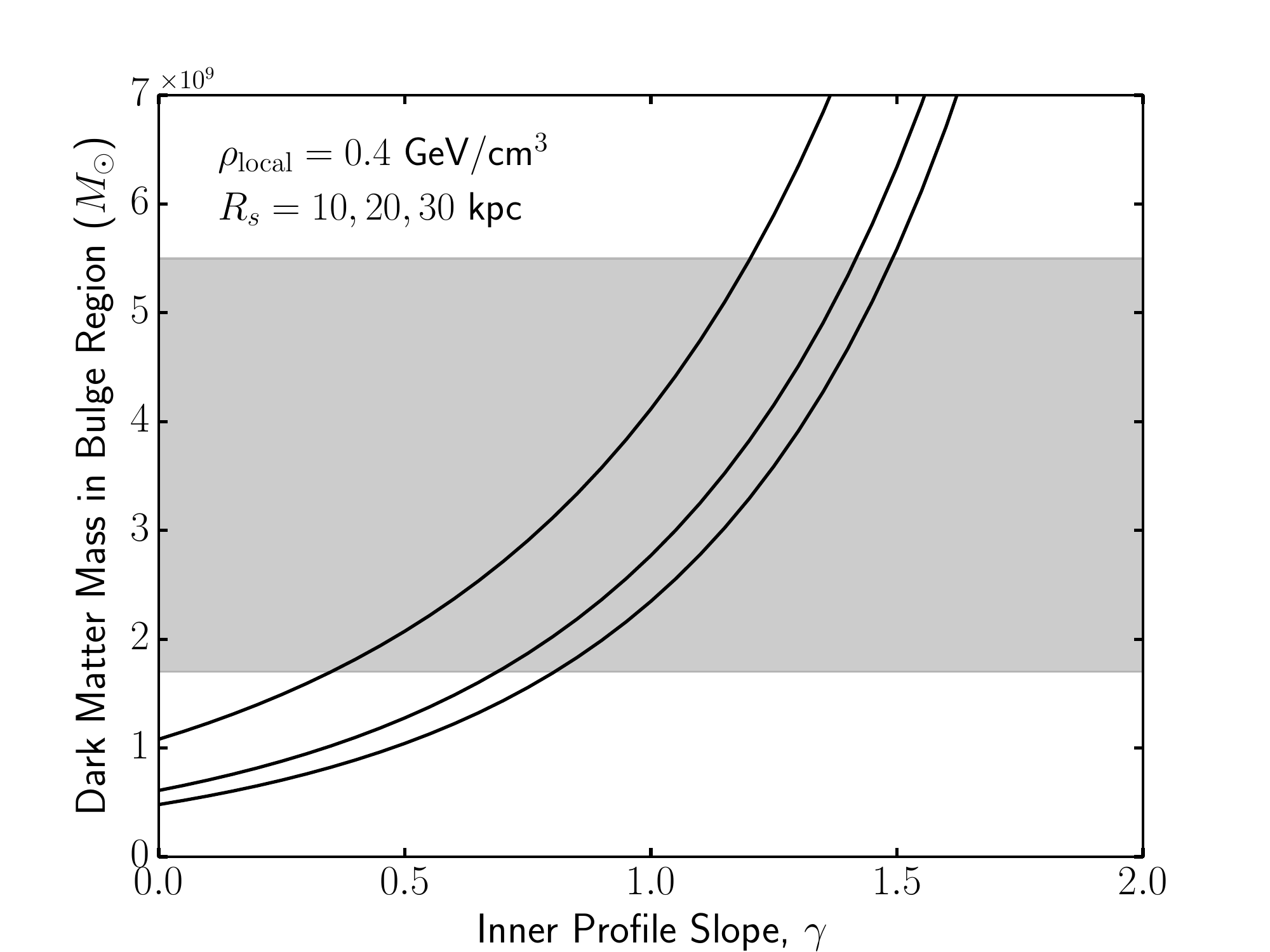}
\caption{The total mass in dark matter within the bulge-bar region of the Milky Way (defined as a $\pm 2.2 \times \pm 1.4 \times \pm 1.2$ kpc box, centered around the Galactic Center), as a function of the inner slope of a generalized NFW halo profile, and for three values of the scale radius (10, 20 and 30 kpc, from left-to-right). We adopt a local dark matter density (at $r$=8.5 kpc) of 0.4 GeV/cm$^3$. The shaded grey region represents the range favored by the determination of  Portail {\it et al.}~\cite{2015MNRAS.448..713P}.}
\label{gamma}
\end{figure}

\begin{figure}
\includegraphics[keepaspectratio,width=0.34\textwidth]{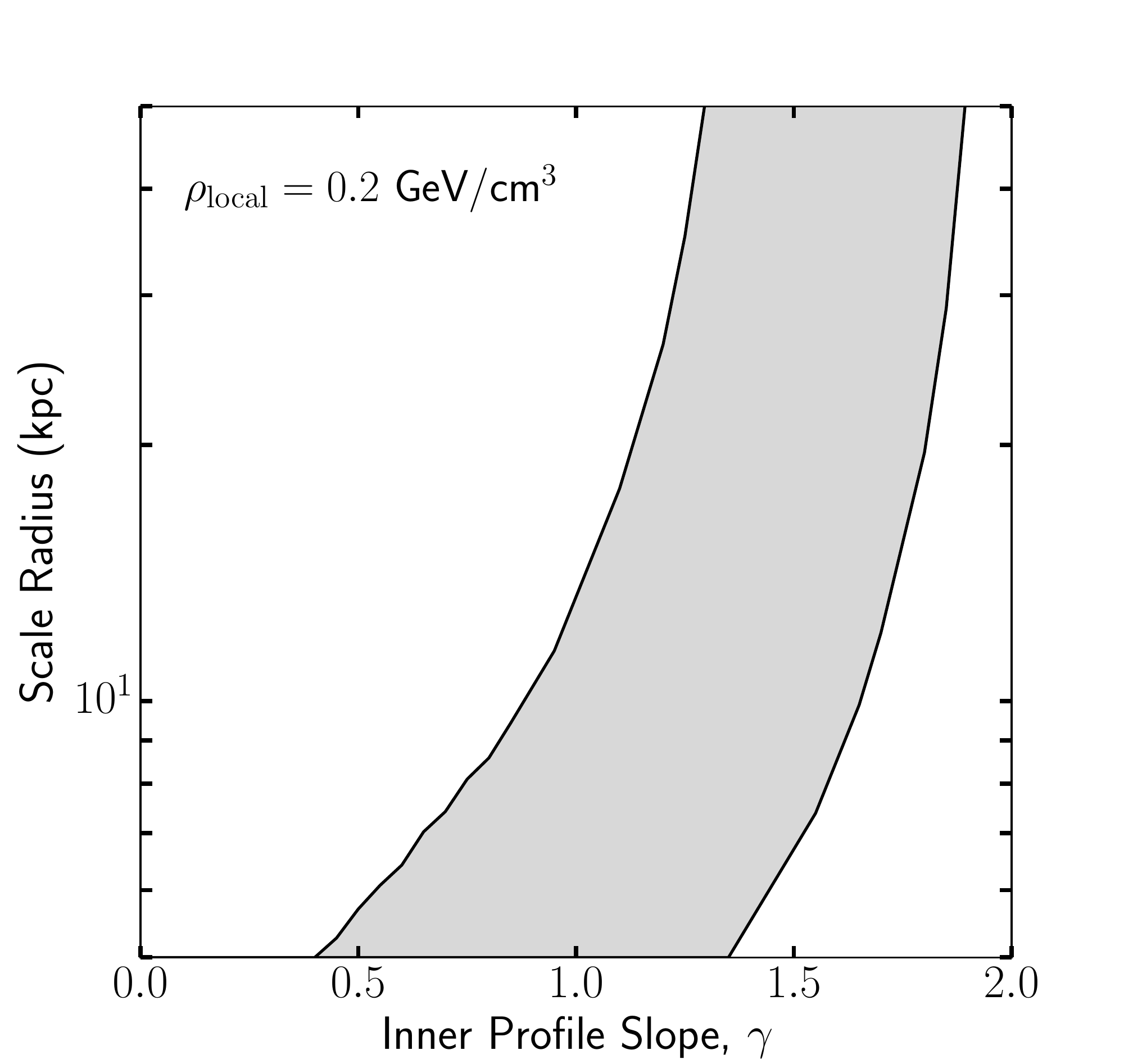}
\hspace{-0.5cm}
\includegraphics[keepaspectratio,width=0.34\textwidth]{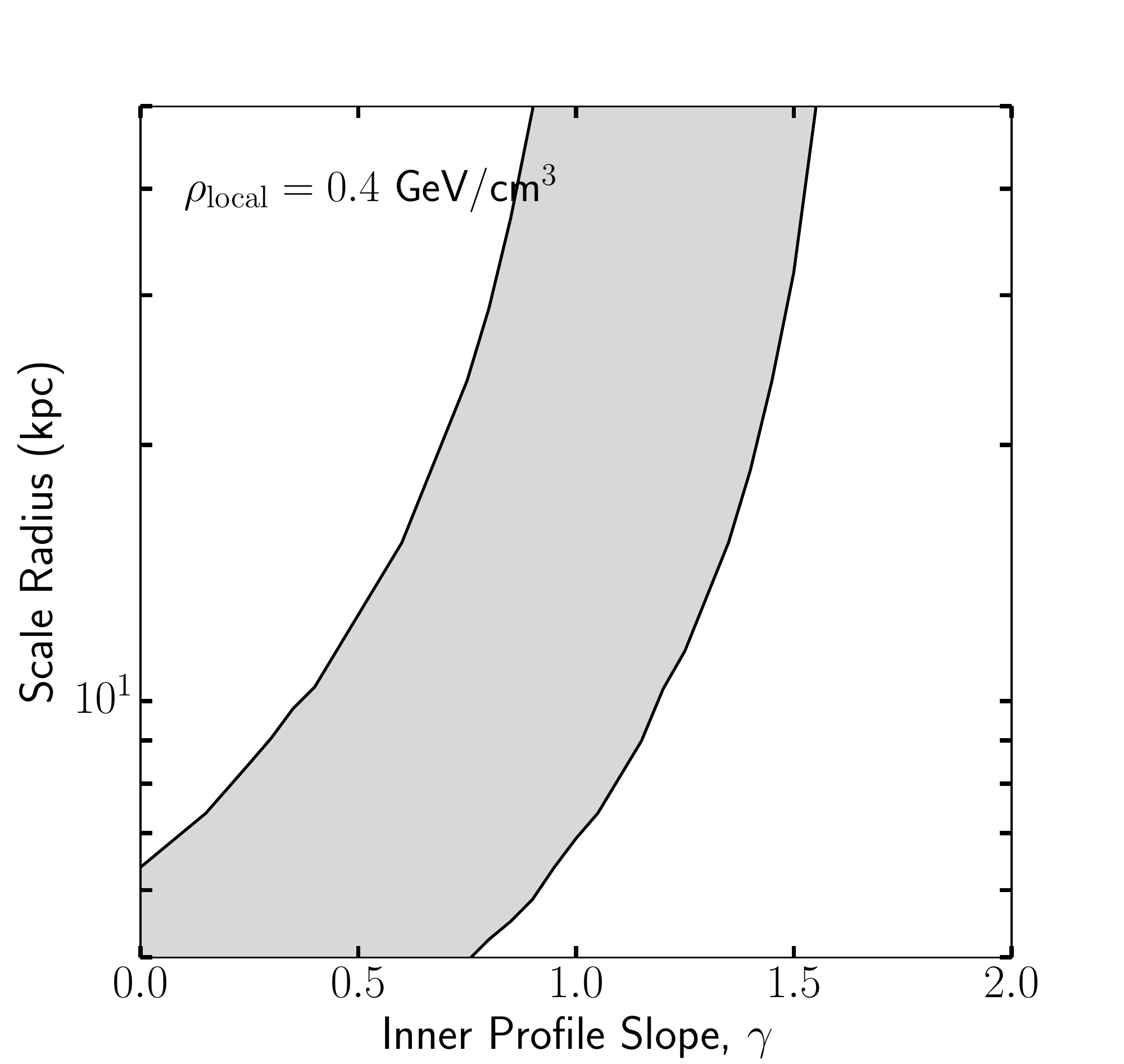}
\hspace{-0.5cm}
\includegraphics[keepaspectratio,width=0.34\textwidth]{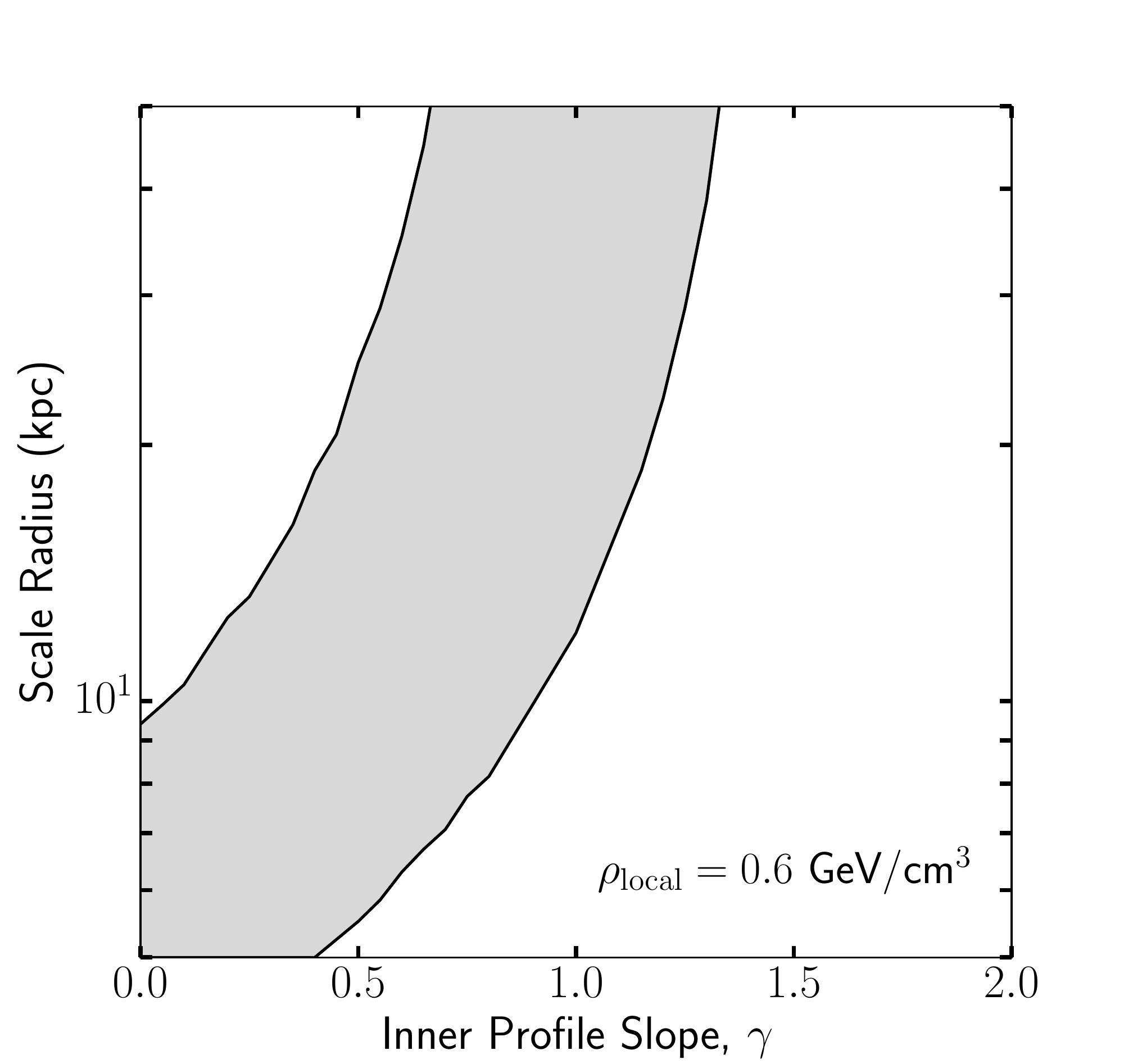}
\caption{Values of the scale radius and inner slope which correspond to a quantity of dark matter in the bulge-bar region that falls within the range favored by Portail {\it et al.}~\cite{2015MNRAS.448..713P}, for three values of the local dark matter density.}
\label{contours}
\end{figure}

In Fig.~\ref{gamma}, we plot the total mass in dark matter within the bulge-bar region as defined by Portail {\it et al.} (a $\pm 2.2 \times \pm 1.4 \times \pm 1.2$ kpc rectangular box, centered around the Galactic Center), as a function of the inner slope of the halo profile, and for three values of the scale radius. In this figure, we adopt a local dark matter density (at $r$=8.5 kpc) of 0.4 GeV/cm$^3$. The shaded grey region of this figure represents the range of values favored by Portail {\it et al.} ($1.7\times 10^9 \, M_{\odot}$ to $5.3\times 10^9 \, M_{\odot}$)~\cite{2015MNRAS.448..713P}. From this figure, it is clear that cusped halo profiles, of the type long favored by numerical simulations of cold collisions dark matter~\cite{Navarro:2008kc,Diemand:2008in}, are in good agreement with the results of Ref.~\cite{2015MNRAS.448..713P}. For example, for $R_s=20$ kpc and $\rho_{\rm local}=0.4$ GeV/cm$^3$, Portail {\it et al.} favor an inner slope of 0.69 to 1.40, approximately centered around the standard NFW value of $\gamma=1.0$. In Fig.~\ref{contours}, we show the values of the scale radius and inner slope that fall within the range favored by Portail {\it et al.}, for three values of the local dark matter density. We also note that the results of Portail {\it et al.} appear to modestly disfavor dark matter distributions that follow an Einasto profile~\cite{Springel:2008cc}. For example, an Einasto profile with $\alpha=0.17$, $\rho_{\rm local}=0.4$ GeV/cm$^3$ and $R_s=20$ kpc, yields a dark matter mass for the bulge-bar region of $0.90\times 10^{9} \, M_{\odot}$, well below the range favored by Portail {\it et al.}

\begin{figure}
\includegraphics[keepaspectratio,width=0.7\textwidth]{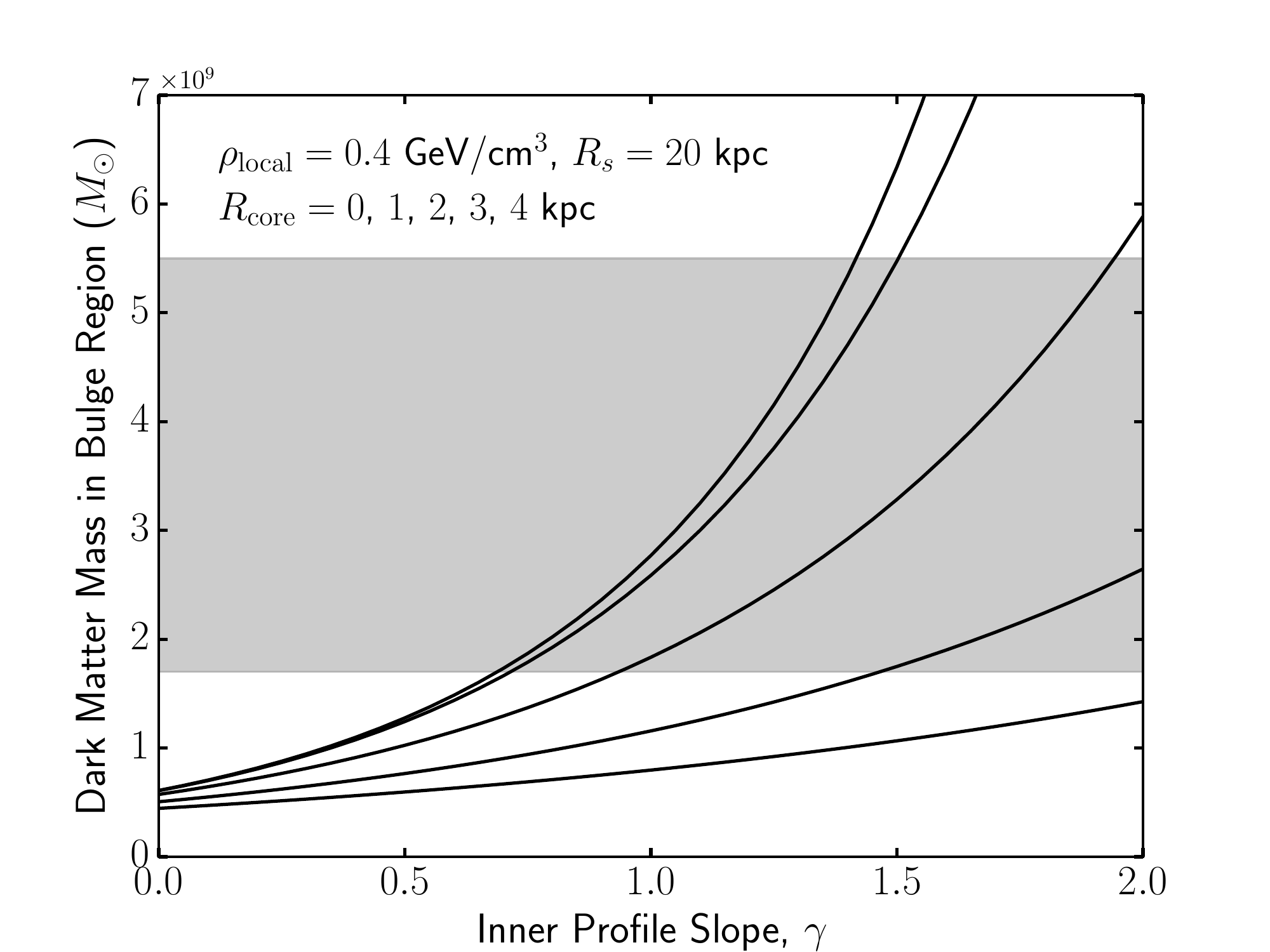}
\caption{As in Fig.~\ref{gamma}, but for profiles with a constant density core. Profiles with a core that is larger than a few kpc in radius are disfavored by the results of Portail {\it et al.}~\cite{2015MNRAS.448..713P}. }
\label{cores}
\end{figure}

Despite the fact that numerical simulations of cold, collisionless dark matter favor cusped halo profiles~\cite{Navarro:2008kc,Diemand:2008in}, such simulations often do not include the potentially important effects of baryons. And although hydrodynamical simulations of Milky Way like systems have improved significantly in recent years, a consensus has not yet emerged regarding whether baryonic effects are more likely to steepen or flatten the dark matter density profile in the innermost kiloparsecs of the Galaxy~\cite{Gnedin:2011uj,Gnedin:2004cx,Governato:2012fa,Kuhlen:2012qw,Weinberg:2001gm,Weinberg:2006ps,Sellwood:2002vb,Valenzuela:2002np,Colin:2005rr,Scannapieco:2011yd,Calore:2015oya,Schaller:2014uwa,DiCintio:2014xia,DiCintio:2013qxa,Schaller:2015mua,Bernal:2016guq}. With this in mind, we show in Fig.~\ref{cores} results for dark matter halos with a constant density core. From this figure, we see that any profile with a core larger than a few kpc in radius is disfavored by the results of Portail {\it et al.}

\begin{figure}
\includegraphics[keepaspectratio,width=0.7\textwidth]{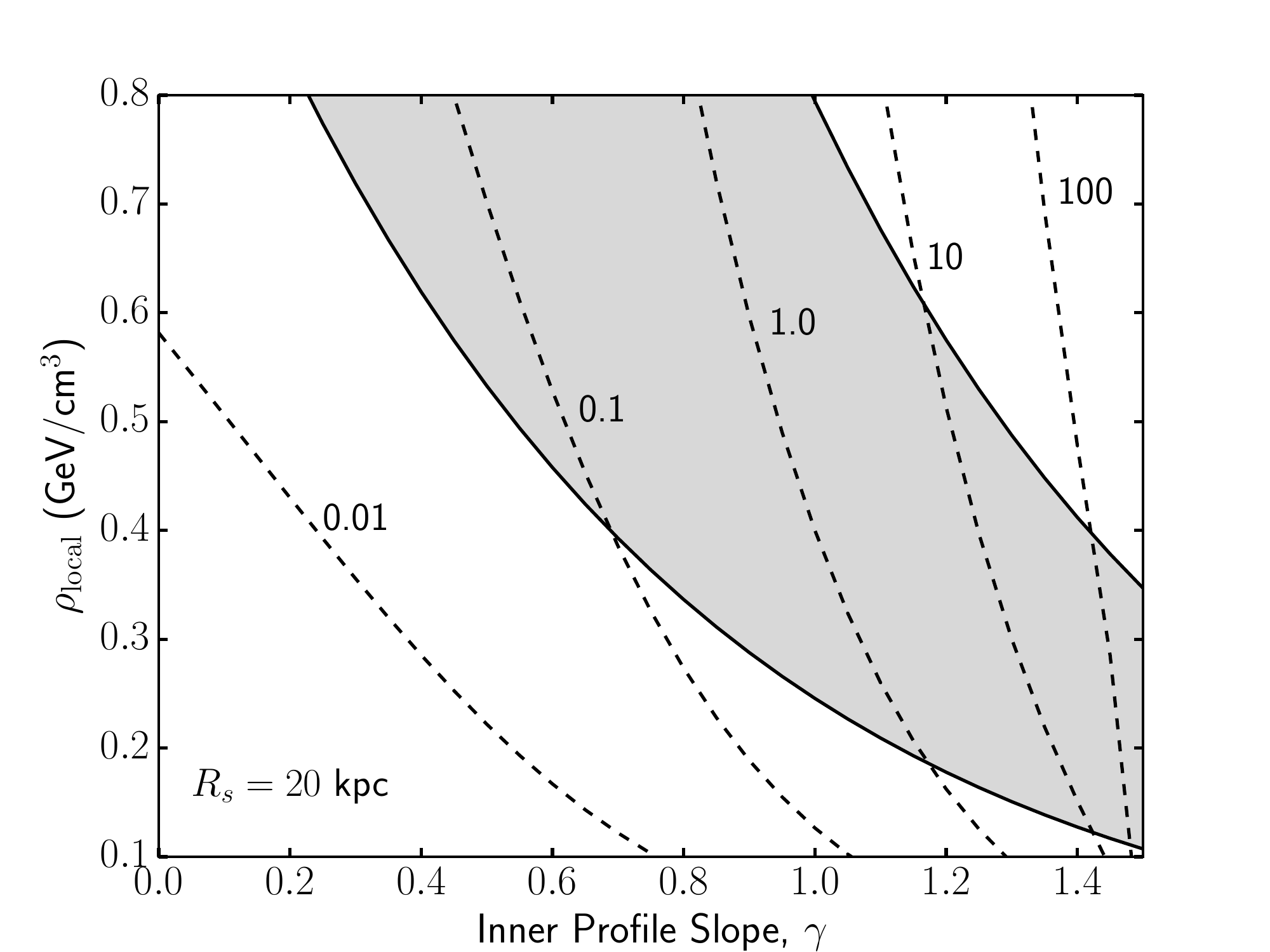}
\caption{The shaded grey region in this figure represents the values of the local density and the inner slope that are favored by the determination of Portail {\it et al.}~\cite{2015MNRAS.448..713P}, for a scale radius of 20 kpc. The dashed lines represent contours of constant $J$-factor, evaluated over the a 1$^{\circ}$ radius around the Galactic Center, with values equal to 0.01, 0.1, 1.0, 10.0 and 100 times that obtained for the case of a standard NFW profile ($\gamma=1$) with $\rho_{\rm local}=0.4$ GeV/cm$^3$.}
\label{jfactor}
\end{figure}

In Fig.~\ref{jfactor}, we show the range of values of the local dark matter density and the inner profile slope that are favored by the results of Portail {\it et al.}, for a scale radius of 20 kpc. As dashed lines, we show contours of constant $J$-factor, evaluated over the a 1$^{\circ}$ radius around the Galactic Center, normalized such that a value of one corresponds to the case of $\gamma=1$ and $\rho_{\rm local}=0.4$ GeV/cm$^3$.  Over most of the parameter space favored by Portail {\it et al.}, we find $J$-factors that lie within approximately an order of magnitude of the value predicted in the standard NFW case.

Analyses of Fermi data have identified a bright and statistically significant excess of GeV-scale gamma-rays from the region surrounding the Galactic Center~\cite{Goodenough:2009gk,Hooper:2010mq,Hooper:2011ti,Abazajian:2012pn,Gordon:2013vta,Hooper:2013rwa,Daylan:2014rsa,Calore:2014xka,TheFermi-LAT:2015kwa,Carlson:2016iis}, with morphological and spectral features consistent with those predicted from annihilating dark matter. To fit the data with annihilating dark matter, these analyses generally find that a halo profile with an inner slope of $\gamma \simeq 1.0-1.4$ is required~\cite{Abazajian:2012pn,Gordon:2013vta,Hooper:2013rwa,Daylan:2014rsa,Calore:2014xka,TheFermi-LAT:2015kwa,Carlson:2016iis}. For the best-fit value of $\gamma=1.28$, as found in Ref.~\cite{Calore:2014xka} (for $R_s=20$ kpc), the results of Portail {\it et al.} favor a local density in the range of 0.156 to 0.504 GeV/cm$^3$. This range of halo normalizations corresponds to annihilation cross sections of $\sigma v = (0.25-2.6)\times 10^{-26}$ cm$^3$/s, for the best-fit mass of $m_{\rm DM}=49$ GeV (annihilating to $b\bar{b}$). Note that this range does not include uncertainties in the determination of the intensity or inner slope.

Lastly, we point out that the $J$-factor for the Galactic Center could be larger than those inferred in this paper if there exists a density spike in the immediate vicinity around the Galaxy's supermassive black hole~\cite{Gondolo:1999ef,Fields:2014pia}. Such a feature could plausibly enhance the flux of annihilation products from the direction of the Galactic Center, but without impacting the flux from other directions of the sky.

\section{Summary and Conclusions}

In this paper, we have made use of a recent dynamical determination of the dark matter mass in the Galactic Bulge-Bar region~\cite{2015MNRAS.448..713P} to constrain the Milky Way's dark matter density profile. We find that these results favor a cusped distribution, consistent with commonly used profile models. For example, for a halo with local density of 0.4 GeV/cm$^3$ and a scale radius of 20 kpc, we find that these results favor an inner slope in the range of $\gamma=$~0.69 to 1.40, consistent with the standard NFW value of $\gamma=1.0$. Profiles with large flat-density cores are disfavored by this determination. Dark matter scenarios capable of accounting for the GeV excess observed from the Inner Galaxy require profiles with $\gamma \simeq 1.0-1.4$, in good agreement with the results presented here.

Our ability to constrain the Milky Way's dark matter halo profile is expected to improve very significantly in the future. In particular, we expect observations from the Wide-Field Infrared Survey Telescope (WFIRST) to improve upon existing dynamical measurements of the Inner Galaxy by an order-of-magnitude~\cite{2015JKAS...48...93G,2015arXiv150303757S}. Combining such observations with those from Gaia and LSST will make it possible to constrain the Milky Way's dark matter halo profile over a wide range of scales, and with much greater precision than is currently possible.

\bigskip
\bigskip

\textbf{Acknowledgments.} We would like to thank David Nataf for extremely helpful discussions, as well as for bringing the work of Portail {\it et al.} to our attention. DH is supported by the US Department of Energy under contract DE-FG02-13ER41958. Fermilab is operated by Fermi Research Alliance, LLC, under Contract No. DE-AC02-07CH11359 with the US Department of Energy.

\bibliography{dminbulge}

\providecommand{\href}[2]{#2}\begingroup\raggedright\begin{thebibliography}{10}

\bibitem{Hooper:2012sr}
D.~Hooper, C.~Kelso and F.~S. Queiroz, \emph{{Stringent and Robust Constraints
  on the Dark Matter Annihilation Cross Section From the Region of the Galactic
  Center}},
  \href{http://dx.doi.org/10.1016/j.astropartphys.2013.04.007}{\emph{Astropart.
  Phys.} {\bf 46} (2013) 55--70}, [\href{http://arxiv.org/abs/1209.3015}{{\tt
  1209.3015}}].

\bibitem{Gomez-Vargas:2013bea}
G.~A. Gómez-Vargas, M.~A. Sánchez-Conde, J.-H. Huh, M.~Peiró, F.~Prada,
  A.~Morselli et~al., \emph{{Constraints on WIMP annihilation for contracted
  dark matterin the inner Galaxy with the Fermi-LAT}},
  \href{http://dx.doi.org/10.1088/1475-7516/2013/10/029}{\emph{JCAP} {\bf 1310}
  (2013) 029}, [\href{http://arxiv.org/abs/1308.3515}{{\tt 1308.3515}}].

\bibitem{Abramowski:2011hc}
{\scshape HESS} collaboration, A.~Abramowski et~al., \emph{{Search for a Dark
  Matter annihilation signal from the Galactic Center halo with H.E.S.S}},
  \href{http://dx.doi.org/10.1103/PhysRevLett.106.161301}{\emph{Phys. Rev.
  Lett.} {\bf 106} (2011) 161301}, [\href{http://arxiv.org/abs/1103.3266}{{\tt
  1103.3266}}].

\bibitem{Aharonian:2006wh}
{\scshape HESS} collaboration, F.~Aharonian et~al., \emph{{H.E.S.S.
  observations of the Galactic Center region and their possible dark matter
  interpretation}},
  \href{http://dx.doi.org/10.1103/PhysRevLett.97.221102}{\emph{Phys. Rev.
  Lett.} {\bf 97} (2006) 221102},
  [\href{http://arxiv.org/abs/astro-ph/0610509}{{\tt astro-ph/0610509}}].

\bibitem{Silverwood:2014yza}
H.~Silverwood, C.~Weniger, P.~Scott and G.~Bertone, \emph{{A realistic
  assessment of the CTA sensitivity to dark matter annihilation}},
  \href{http://dx.doi.org/10.1088/1475-7516/2015/03/055}{\emph{JCAP} {\bf 1503}
  (2015) 055}, [\href{http://arxiv.org/abs/1408.4131}{{\tt 1408.4131}}].

\bibitem{Pierre:2014tra}
M.~Pierre, J.~M. Siegal-Gaskins and P.~Scott, \emph{{Sensitivity of CTA to dark
  matter signals from the Galactic Center}},
  \href{http://dx.doi.org/10.1088/1475-7516/2014/10/E01,
  10.1088/1475-7516/2014/06/024}{\emph{JCAP} {\bf 1406} (2014) 024},
  [\href{http://arxiv.org/abs/1401.7330}{{\tt 1401.7330}}].

\bibitem{Weber:2009pt}
M.~Weber and W.~de~Boer, \emph{{Determination of the Local Dark Matter Density
  in our Galaxy}},
  \href{http://dx.doi.org/10.1051/0004-6361/200913381}{\emph{Astron.
  Astrophys.} {\bf 509} (2010) A25},
  [\href{http://arxiv.org/abs/0910.4272}{{\tt 0910.4272}}].

\bibitem{2010JCAP...08..004C}
R.~{Catena} and P.~{Ullio}, \emph{{A novel determination of the local dark
  matter density}},
  \href{http://dx.doi.org/10.1088/1475-7516/2010/08/004}{\emph{JCAP} {\bf 8}
  (Aug., 2010) 4}, [\href{http://arxiv.org/abs/0907.0018}{{\tt 0907.0018}}].

\bibitem{2011JCAP...11..029I}
F.~{Iocco}, M.~{Pato}, G.~{Bertone} and P.~{Jetzer}, \emph{{Dark Matter
  distribution in the Milky Way: microlensing and dynamical constraints}},
  \href{http://dx.doi.org/10.1088/1475-7516/2011/11/029}{\emph{JCAP} {\bf 11}
  (Nov., 2011) 29}, [\href{http://arxiv.org/abs/1107.5810}{{\tt 1107.5810}}].

\bibitem{Bovy:2012tw}
J.~Bovy and S.~Tremaine, \emph{{On the local dark matter density}},
  \href{http://dx.doi.org/10.1088/0004-637X/756/1/89}{\emph{Astrophys. J.} {\bf
  756} (2012) 89}, [\href{http://arxiv.org/abs/1205.4033}{{\tt 1205.4033}}].

\bibitem{Garbari:2012ff}
S.~Garbari, C.~Liu, J.~I. Read and G.~Lake, \emph{{A new determination of the
  local dark matter density from the kinematics of K dwarfs}},
  \href{http://dx.doi.org/10.1111/j.1365-2966.2012.21608.x}{\emph{Mon. Not.
  Roy. Astron. Soc.} {\bf 425} (2012) 1445},
  [\href{http://arxiv.org/abs/1206.0015}{{\tt 1206.0015}}].

\bibitem{Bovy:2013raa}
J.~Bovy and H.-W. Rix, \emph{{A Direct Dynamical Measurement of the Milky Way's
  Disk Surface Density Profile, Disk Scale Length, and Dark Matter Profile at 4
  kpc $\stackrel{<}{\sim}$ R $\stackrel{<}{\sim}$ 9 kpc}},
  \href{http://dx.doi.org/10.1088/0004-637X/779/2/115}{\emph{Astrophys. J.}
  {\bf 779} (2013) 115}, [\href{http://arxiv.org/abs/1309.0809}{{\tt
  1309.0809}}].

\bibitem{Read:2014qva}
J.~I. Read, \emph{{The Local Dark Matter Density}},
  \href{http://dx.doi.org/10.1088/0954-3899/41/6/063101}{\emph{J. Phys.} {\bf
  G41} (2014) 063101}, [\href{http://arxiv.org/abs/1404.1938}{{\tt
  1404.1938}}].

\bibitem{Iocco:2015xga}
F.~Iocco, M.~Pato and G.~Bertone, \emph{{Evidence for dark matter in the inner
  Milky Way}}, \href{http://dx.doi.org/10.1038/nphys3237}{\emph{Nature Phys.}
  {\bf 11} (2015) 245--248}, [\href{http://arxiv.org/abs/1502.03821}{{\tt
  1502.03821}}].

\bibitem{Pato:2015dua}
M.~Pato, F.~Iocco and G.~Bertone, \emph{{Dynamical constraints on the dark
  matter distribution in the Milky Way}},
  \href{http://dx.doi.org/10.1088/1475-7516/2015/12/001}{\emph{JCAP} {\bf 1512}
  (2015) 001}, [\href{http://arxiv.org/abs/1504.06324}{{\tt 1504.06324}}].

\bibitem{2015MNRAS.448..713P}
M.~{Portail}, C.~{Wegg}, O.~{Gerhard} and I.~{Martinez-Valpuesta},
  \emph{{Made-to-measure models of the Galactic box/peanut bulge: stellar and
  total mass in the bulge region}},
  \href{http://dx.doi.org/10.1093/mnras/stv058}{\emph{\mnras} {\bf 448} (Mar.,
  2015) 713--731}, [\href{http://arxiv.org/abs/1502.00633}{{\tt 1502.00633}}].

\bibitem{2013MNRAS.435.1874W}
C.~{Wegg} and O.~{Gerhard}, \emph{{Mapping the three-dimensional density of the
  Galactic bulge with VVV red clump stars}},
  \href{http://dx.doi.org/10.1093/mnras/stt1376}{\emph{\mnras} {\bf 435} (Nov.,
  2013) 1874--1887}, [\href{http://arxiv.org/abs/1308.0593}{{\tt 1308.0593}}].

\bibitem{2015MNRAS.450.4050W}
C.~{Wegg}, O.~{Gerhard} and M.~{Portail}, \emph{{The structure of the Milky
  Way's bar outside the bulge}},
  \href{http://dx.doi.org/10.1093/mnras/stv745}{\emph{\mnras} {\bf 450} (July,
  2015) 4050--4069}.

\bibitem{2008ApJ...688.1060H}
C.~D. {Howard}, R.~M. {Rich}, D.~B. {Reitzel}, A.~{Koch}, R.~{De Propris} and
  H.~{Zhao}, \emph{{The Bulge Radial Velocity Assay (BRAVA). I. Sample
  Selection and a Rotation Curve}},
  \href{http://dx.doi.org/10.1086/592106}{\emph{\apj} {\bf 688} (Dec., 2008)
  1060--1077}, [\href{http://arxiv.org/abs/0807.3967}{{\tt 0807.3967}}].

\bibitem{2012AJ....143...57K}
A.~{Kunder}, A.~{Koch}, R.~M. {Rich}, R.~{de Propris}, C.~D. {Howard}, S.~A.
  {Stubbs} et~al., \emph{{The Bulge Radial Velocity Assay (BRAVA). II. Complete
  Sample and Data Release}},
  \href{http://dx.doi.org/10.1088/0004-6256/143/3/57}{\emph{\aj} {\bf 143}
  (Mar., 2012) 57}, [\href{http://arxiv.org/abs/1112.1955}{{\tt 1112.1955}}].

\bibitem{1992ApJ...387..181K}
S.~M. {Kent}, \emph{{Galactic structure from the spacelab infrared telescope.
  III - A dynamical model for the Milky Way bulge}},
  \href{http://dx.doi.org/10.1086/171070}{\emph{\apj} {\bf 387} (Mar., 1992)
  181--188}.

\bibitem{2003MNRAS.340..949B}
N.~{Bissantz}, P.~{Englmaier} and O.~{Gerhard}, \emph{{Gas dynamics in the
  Milky Way: second pattern speed and large-scale morphology}},
  \href{http://dx.doi.org/10.1046/j.1365-8711.2003.06358.x}{\emph{\mnras} {\bf
  340} (Apr., 2003) 949--968},
  [\href{http://arxiv.org/abs/astro-ph/0212516}{{\tt astro-ph/0212516}}].

\bibitem{1995ApJ...449..521H}
C.~{Han} and A.~{Gould}, \emph{{Statistics of Microlensing Optical Depth}},
  \href{http://dx.doi.org/10.1086/176076}{\emph{\apj} {\bf 449} (Aug., 1995)
  521}, [\href{http://arxiv.org/abs/astro-ph/9410052}{{\tt astro-ph/9410052}}].

\bibitem{1995ApJ...444L..89B}
R.~D. {Blum}, \emph{{Figure rotation and the mass of the galactic bulge}},
  \href{http://dx.doi.org/10.1086/187867}{\emph{\apjl} {\bf 444} (May, 1995)
  L89--L91}, [\href{http://arxiv.org/abs/astro-ph/9503046}{{\tt
  astro-ph/9503046}}].

\bibitem{2015MNRAS.447.1535N}
D.~M. {Nataf}, A.~{Udalski}, J.~{Skowron}, M.~K. {Szyma{\'n}ski}, M.~{Kubiak},
  G.~{Pietrzy{\'n}ski} et~al., \emph{{The X-shaped Milky Way bulge in OGLE-III
  photometry and in N-body models}},
  \href{http://dx.doi.org/10.1093/mnras/stu2497}{\emph{\mnras} {\bf 447} (Feb.,
  2015) 1535--1549}, [\href{http://arxiv.org/abs/1411.7384}{{\tt 1411.7384}}].

\bibitem{2013ApJ...769...88N}
D.~M. {Nataf}, A.~{Gould}, P.~{Fouqu{\'e}}, O.~A. {Gonzalez}, J.~A. {Johnson},
  J.~{Skowron} et~al., \emph{{Reddening and Extinction toward the Galactic
  Bulge from OGLE-III: The Inner Milky Way's R$_{V}$ \~{} 2.5 Extinction
  Curve}}, \href{http://dx.doi.org/10.1088/0004-637X/769/2/88}{\emph{\apj} {\bf
  769} (June, 2013) 88}, [\href{http://arxiv.org/abs/1208.1263}{{\tt
  1208.1263}}].

\bibitem{Navarro:1995iw}
J.~F. Navarro, C.~S. Frenk and S.~D. White, \emph{{The Structure of cold dark
  matter halos}}, \href{http://dx.doi.org/10.1086/177173}{\emph{Astrophys.J.}
  {\bf 462} (1996) 563--575},
  [\href{http://arxiv.org/abs/astro-ph/9508025}{{\tt astro-ph/9508025}}].

\bibitem{Navarro:1996gj}
J.~F. Navarro, C.~S. Frenk and S.~D. White, \emph{{A Universal density profile
  from hierarchical clustering}},
  \href{http://dx.doi.org/10.1086/304888}{\emph{Astrophys.J.} {\bf 490} (1997)
  493--508}, [\href{http://arxiv.org/abs/astro-ph/9611107}{{\tt
  astro-ph/9611107}}].

\bibitem{Bergstrom:1997fj}
L.~Bergstrom, P.~Ullio and J.~H. Buckley, \emph{{Observability of gamma-rays
  from dark matter neutralino annihilations in the Milky Way halo}},
  \href{http://dx.doi.org/10.1016/S0927-6505(98)00015-2}{\emph{Astropart.
  Phys.} {\bf 9} (1998) 137--162},
  [\href{http://arxiv.org/abs/astro-ph/9712318}{{\tt astro-ph/9712318}}].

\bibitem{Navarro:2008kc}
J.~F. Navarro, A.~Ludlow, V.~Springel, J.~Wang, M.~Vogelsberger, S.~D.~M. White
  et~al., \emph{{The Diversity and Similarity of Cold Dark Matter Halos}},
  \href{http://dx.doi.org/10.1111/j.1365-2966.2009.15878.x}{\emph{Mon. Not.
  Roy. Astron. Soc.} {\bf 402} (2010) 21},
  [\href{http://arxiv.org/abs/0810.1522}{{\tt 0810.1522}}].

\bibitem{Diemand:2008in}
J.~Diemand, M.~Kuhlen, P.~Madau, M.~Zemp, B.~Moore et~al., \emph{{Clumps and
  streams in the local dark matter distribution}},
  \href{http://dx.doi.org/10.1038/nature07153}{\emph{Nature} {\bf 454} (2008)
  735--738}, [\href{http://arxiv.org/abs/0805.1244}{{\tt 0805.1244}}].

\bibitem{Springel:2008cc}
V.~Springel, J.~Wang, M.~Vogelsberger, A.~Ludlow, A.~Jenkins et~al., \emph{{The
  Aquarius Project: the subhalos of galactic halos}},
  \href{http://dx.doi.org/10.1111/j.1365-2966.2008.14066.x}{\emph{Mon.Not.Roy.Astron.Soc.}
  {\bf 391} (2008) 1685--1711}, [\href{http://arxiv.org/abs/0809.0898}{{\tt
  0809.0898}}].

\bibitem{Gnedin:2011uj}
O.~Y. Gnedin, D.~Ceverino, N.~Y. Gnedin, A.~A. Klypin, A.~V. Kravtsov,
  R.~Levine et~al., \emph{{Halo Contraction Effect in Hydrodynamic Simulations
  of Galaxy Formation}},  \href{http://arxiv.org/abs/1108.5736}{{\tt
  1108.5736}}.

\bibitem{Gnedin:2004cx}
O.~Y. Gnedin, A.~V. Kravtsov, A.~A. Klypin and D.~Nagai, \emph{{Response of
  dark matter halos to condensation of baryons: Cosmological simulations and
  improved adiabatic contraction model}},
  \href{http://dx.doi.org/10.1086/424914}{\emph{Astrophys. J.} {\bf 616} (2004)
  16--26}, [\href{http://arxiv.org/abs/astro-ph/0406247}{{\tt
  astro-ph/0406247}}].

\bibitem{Governato:2012fa}
F.~Governato, A.~Zolotov, A.~Pontzen, C.~Christensen, S.~H. Oh, A.~M. Brooks
  et~al., \emph{{Cuspy No More: How Outflows Affect the Central Dark Matter and
  Baryon Distribution in Lambda CDM Galaxies}},
  \href{http://dx.doi.org/10.1111/j.1365-2966.2012.20696.x}{\emph{Mon. Not.
  Roy. Astron. Soc.} {\bf 422} (2012) 1231--1240},
  [\href{http://arxiv.org/abs/1202.0554}{{\tt 1202.0554}}].

\bibitem{Kuhlen:2012qw}
M.~Kuhlen, J.~Guedes, A.~Pillepich, P.~Madau and L.~Mayer, \emph{{An Off-center
  Density Peak in the Milky Way's Dark Matter Halo?}},
  \href{http://dx.doi.org/10.1088/0004-637X/765/1/10}{\emph{Astrophys. J.} {\bf
  765} (2013) 10}, [\href{http://arxiv.org/abs/1208.4844}{{\tt 1208.4844}}].

\bibitem{Weinberg:2001gm}
M.~D. Weinberg and N.~Katz, \emph{{Bar-driven dark halo evolution: a resolution
  of the cusp-core controversy}},
  \href{http://dx.doi.org/10.1086/343847}{\emph{Astrophys. J.} {\bf 580} (2002)
  627--633}, [\href{http://arxiv.org/abs/astro-ph/0110632}{{\tt
  astro-ph/0110632}}].

\bibitem{Weinberg:2006ps}
M.~D. Weinberg and N.~Katz, \emph{{The bar-halo interaction. 2. secular
  evolution and the religion of n-body simulations}},
  \href{http://dx.doi.org/10.1111/j.1365-2966.2006.11307.x}{\emph{Mon. Not.
  Roy. Astron. Soc.} {\bf 375} (2007) 460--476},
  [\href{http://arxiv.org/abs/astro-ph/0601138}{{\tt astro-ph/0601138}}].

\bibitem{Sellwood:2002vb}
J.~A. Sellwood, \emph{{Bars and dark matter halo cores}},
  \href{http://dx.doi.org/10.1086/368285}{\emph{Astrophys. J.} {\bf 587} (2003)
  638--648}, [\href{http://arxiv.org/abs/astro-ph/0210079}{{\tt
  astro-ph/0210079}}].

\bibitem{Valenzuela:2002np}
O.~Valenzuela and A.~Klypin, \emph{{Secular bar formation in galaxies with
  significant amount of dark matter}},
  \href{http://dx.doi.org/10.1046/j.1365-8711.2003.06947.x}{\emph{Mon. Not.
  Roy. Astron. Soc.} {\bf 345} (2003) 406},
  [\href{http://arxiv.org/abs/astro-ph/0204028}{{\tt astro-ph/0204028}}].

\bibitem{Colin:2005rr}
P.~Colin, O.~Valenzuela and A.~Klypin, \emph{{Bars and cold dark matter
  halos}}, \href{http://dx.doi.org/10.1086/503791}{\emph{Astrophys. J.} {\bf
  644} (2006) 687--700}, [\href{http://arxiv.org/abs/astro-ph/0506627}{{\tt
  astro-ph/0506627}}].

\bibitem{Scannapieco:2011yd}
C.~Scannapieco et~al., \emph{{The Aquila comparison Project: The Effects of
  Feedback and Numerical Methods on Simulations of Galaxy Formation}},
  \href{http://dx.doi.org/10.1111/j.1365-2966.2012.20993.x}{\emph{Mon. Not.
  Roy. Astron. Soc.} {\bf 423} (2012) 1726},
  [\href{http://arxiv.org/abs/1112.0315}{{\tt 1112.0315}}].

\bibitem{Calore:2015oya}
F.~Calore, N.~Bozorgnia, M.~Lovell, G.~Bertone, M.~Schaller, C.~S. Frenk
  et~al., \emph{{Simulated Milky Way analogues: implications for dark matter
  indirect searches}},
  \href{http://dx.doi.org/10.1088/1475-7516/2015/12/053}{\emph{JCAP} {\bf 1512}
  (2015) 053}, [\href{http://arxiv.org/abs/1509.02164}{{\tt 1509.02164}}].

\bibitem{Schaller:2014uwa}
M.~Schaller, C.~S. Frenk, R.~G. Bower, T.~Theuns, A.~Jenkins, J.~Schaye et~al.,
  \emph{{Baryon effects on the internal structure of ΛCDM haloes in the EAGLE
  simulations}}, \href{http://dx.doi.org/10.1093/mnras/stv1067}{\emph{Mon. Not.
  Roy. Astron. Soc.} {\bf 451} (2015) 1247--1267},
  [\href{http://arxiv.org/abs/1409.8617}{{\tt 1409.8617}}].

\bibitem{DiCintio:2014xia}
A.~Di~Cintio, C.~B. Brook, A.~A. Dutton, A.~V. Macciò, G.~S. Stinson and
  A.~Knebe, \emph{{A mass-dependent density profile for dark matter haloes
  including the influence of galaxy formation}},
  \href{http://dx.doi.org/10.1093/mnras/stu729}{\emph{Mon. Not. Roy. Astron.
  Soc.} {\bf 441} (2014) 2986--2995},
  [\href{http://arxiv.org/abs/1404.5959}{{\tt 1404.5959}}].

\bibitem{DiCintio:2013qxa}
A.~Di~Cintio, C.~B. Brook, A.~V. Macciò, G.~S. Stinson, A.~Knebe, A.~A. Dutton
  et~al., \emph{{The dependence of dark matter profiles on the stellar-to-halo
  mass ratio: a prediction for cusps versus cores}},
  \href{http://dx.doi.org/10.1093/mnras/stt1891}{\emph{Mon. Not. Roy. Astron.
  Soc.} {\bf 437} (2014) 415--423}, [\href{http://arxiv.org/abs/1306.0898}{{\tt
  1306.0898}}].

\bibitem{Schaller:2015mua}
M.~Schaller et~al., \emph{{Dark matter annihilation radiation in hydrodynamic
  simulations of Milky Way haloes}},
  \href{http://dx.doi.org/10.1093/mnras/stv2667}{\emph{Mon. Not. Roy. Astron.
  Soc.} {\bf 455} (2016) 4442--4451},
  [\href{http://arxiv.org/abs/1509.02166}{{\tt 1509.02166}}].

\bibitem{Bernal:2016guq}
N.~Bernal, L.~Necib and T.~R. Slatyer, \emph{{Spherical Cows in Dark Matter
  Indirect Detection}},  \href{http://arxiv.org/abs/1606.00433}{{\tt
  1606.00433}}.

\bibitem{Goodenough:2009gk}
L.~Goodenough and D.~Hooper, \emph{{Possible Evidence For Dark Matter
  Annihilation In The Inner Milky Way From The Fermi Gamma Ray Space
  Telescope}},  \href{http://arxiv.org/abs/0910.2998}{{\tt 0910.2998}}.

\bibitem{Hooper:2010mq}
D.~Hooper and L.~Goodenough, \emph{{Dark Matter Annihilation in The Galactic
  Center As Seen by the Fermi Gamma Ray Space Telescope}},
  \href{http://dx.doi.org/10.1016/j.physletb.2011.02.029}{\emph{Phys.Lett.}
  {\bf B697} (2011) 412--428}, [\href{http://arxiv.org/abs/1010.2752}{{\tt
  1010.2752}}].

\bibitem{Hooper:2011ti}
D.~Hooper and T.~Linden, \emph{{On The Origin Of The Gamma Rays From The
  Galactic Center}},
  \href{http://dx.doi.org/10.1103/PhysRevD.84.123005}{\emph{Phys.Rev.} {\bf
  D84} (2011) 123005}, [\href{http://arxiv.org/abs/1110.0006}{{\tt
  1110.0006}}].

\bibitem{Abazajian:2012pn}
K.~N. Abazajian and M.~Kaplinghat, \emph{{Detection of a Gamma-Ray Source in
  the Galactic Center Consistent with Extended Emission from Dark Matter
  Annihilation and Concentrated Astrophysical Emission}},
  \href{http://dx.doi.org/10.1103/PhysRevD.86.083511}{\emph{Phys.Rev.} {\bf
  D86} (2012) 083511}, [\href{http://arxiv.org/abs/1207.6047}{{\tt
  1207.6047}}].

\bibitem{Gordon:2013vta}
C.~Gordon and O.~Macias, \emph{{Dark Matter and Pulsar Model Constraints from
  Galactic Center Fermi-LAT Gamma Ray Observations}},
  \href{http://dx.doi.org/10.1103/PhysRevD.88.083521}{\emph{Phys.Rev.} {\bf
  D88} (2013) 083521}, [\href{http://arxiv.org/abs/1306.5725}{{\tt
  1306.5725}}].

\bibitem{Hooper:2013rwa}
D.~Hooper and T.~R. Slatyer, \emph{{Two Emission Mechanisms in the Fermi
  Bubbles: A Possible Signal of Annihilating Dark Matter}},
  \href{http://dx.doi.org/10.1016/j.dark.2013.06.003}{\emph{Phys.Dark Univ.}
  {\bf 2} (2013) 118--138}, [\href{http://arxiv.org/abs/1302.6589}{{\tt
  1302.6589}}].

\bibitem{Daylan:2014rsa}
T.~Daylan, D.~P. Finkbeiner, D.~Hooper, T.~Linden, S.~K.~N. Portillo, N.~L.
  Rodd et~al., \emph{{The characterization of the gamma-ray signal from the
  central Milky Way: A case for annihilating dark matter}},
  \href{http://dx.doi.org/10.1016/j.dark.2015.12.005}{\emph{Phys. Dark Univ.}
  {\bf 12} (2016) 1--23}, [\href{http://arxiv.org/abs/1402.6703}{{\tt
  1402.6703}}].

\bibitem{Calore:2014xka}
F.~Calore, I.~Cholis and C.~Weniger, \emph{{Background model systematics for
  the Fermi GeV excess}},
  \href{http://dx.doi.org/10.1088/1475-7516/2015/03/038}{\emph{JCAP} {\bf 1503}
  (2015) 038}, [\href{http://arxiv.org/abs/1409.0042}{{\tt 1409.0042}}].

\bibitem{TheFermi-LAT:2015kwa}
{\scshape Fermi-LAT} collaboration, M.~Ajello et~al., \emph{{Fermi-LAT
  Observations of High-Energy $\gamma$-Ray Emission Toward the Galactic
  Center}},
  \href{http://dx.doi.org/10.3847/0004-637X/819/1/44}{\emph{Astrophys. J.} {\bf
  819} (2016) 44}, [\href{http://arxiv.org/abs/1511.02938}{{\tt 1511.02938}}].

\bibitem{Carlson:2016iis}
E.~Carlson, T.~Linden and S.~Profumo, \emph{{Improved Cosmic-Ray Injection
  Models and the Galactic Center Gamma-Ray Excess}},
  \href{http://arxiv.org/abs/1603.06584}{{\tt 1603.06584}}.

\bibitem{Gondolo:1999ef}
P.~Gondolo and J.~Silk, \emph{{Dark matter annihilation at the galactic
  center}}, \href{http://dx.doi.org/10.1103/PhysRevLett.83.1719}{\emph{Phys.
  Rev. Lett.} {\bf 83} (1999) 1719--1722},
  [\href{http://arxiv.org/abs/astro-ph/9906391}{{\tt astro-ph/9906391}}].

\bibitem{Fields:2014pia}
B.~D. Fields, S.~L. Shapiro and J.~Shelton, \emph{{Galactic Center Gamma-Ray
  Excess from Dark Matter Annihilation: Is There A Black Hole Spike?}},
  \href{http://dx.doi.org/10.1103/PhysRevLett.113.151302}{\emph{Phys. Rev.
  Lett.} {\bf 113} (2014) 151302}, [\href{http://arxiv.org/abs/1406.4856}{{\tt
  1406.4856}}].

\bibitem{2015JKAS...48...93G}
A.~{Gould}, D.~{Huber}, M.~{Penny} and D.~{Stello}, \emph{{WFIRST Ultra-Precise
  Astrometry II: Asteroseismology}},
  \href{http://dx.doi.org/10.5303/JKAS.2015.48.2.93}{\emph{Journal of Korean
  Astronomical Society} {\bf 48} (Apr., 2015) 93--104},
  [\href{http://arxiv.org/abs/1410.7395}{{\tt 1410.7395}}].

\bibitem{2015arXiv150303757S}
D.~{Spergel}, N.~{Gehrels}, C.~{Baltay}, D.~{Bennett}, J.~{Breckinridge},
  M.~{Donahue} et~al., \emph{{Wide-Field InfrarRed Survey
  Telescope-Astrophysics Focused Telescope Assets WFIRST-AFTA 2015 Report}},
  {\emph{ArXiv e-prints} (Mar., 2015) },
  [\href{http://arxiv.org/abs/1503.03757}{{\tt 1503.03757}}].

\end{thebibliography}\endgroup
\bibliographystyle{JHEP}

\end{document}